\documentclass{amsart}

\theoremstyle{definition}

\theoremstyle{remark}

\numberwithin{equation}{section}

\begin{document}

\title{ Free Energy Adopted Stochastic Optimization of Protein Folding } 

\author{Toshio Fukumi}
\address{Department of Business Administration, Matsuyama University, 2-4 Bunkyo, Matsuyama, Ehime 790 8578, Japan}
\email{tofukumi@cc.matsuyama-u.ac.jp}



\date{\today}



\begin{abstract}
Optimal structure of proteins is described by linear stochastic differential equation with mean decrease of free energy and volatility. Structure determining strategy is given by a twin of stochastic variables for which empirical conditions are not postulated. Optimal structure determination will be deformed to be adoptive to trading strategy employing martingale property where stochastic integral w.r.t. analytical solution of stochastic differential equation  will be employed.
\end{abstract}

\maketitle


\par

\section {Stochastic Optimization}
.
\subsection{Stochastic Differential Equation}
\par
Let us start from heat equation
\begin{equation}
{\partial u\over {\partial t}}={1\over 2} {\partial^2 u^2\over {\partial^2 x^2}},
\end{equation}
the solution is given by
\begin{equation}
 u= {1 \over {\sqrt {2{\pi}t}}}e^  {-{u^2 \over 2t}},
\end{equation}
and corresponding stochastic differential equation is
\begin{equation}
 du_t = dB_t,
\end{equation}
where $B_t$ denotes Brownian motion. In case there is a drift term we have
\begin{equation}
 du_t =  -cdt+dB_t,
\end{equation}
whose solution is
\begin {equation}
 u= {1 \over {\sqrt {2{\pi}t}}}e^  {-{(u-ct)^2 \over 2t}},
\end{equation}
where $c$ is a velocity.
\par
Let us modify the last equation to be
\begin {equation}
 x_t= cdt + {\sqrt \alpha}dB_t,
\end{equation}
where  $\alpha$ is volatility.
This equation can be integrated to be
\begin {equation}
 x_t=\int _0^t{c(t-s)}\sqrt\alpha dB_s .
\end{equation}
Let
\begin {equation}
 X_t=X_0 e^{x_t},.
\end{equation}
and applying Ito formula we obtain
\begin {equation}
d X_t=\mu X_t dt + {\sqrt \alpha}X_tdB_t,
\end{equation}
where
let
\begin {equation}
\mu = c+ \alpha,
\end{equation}
which represents averaged loss of energy,
and this can be integrated to be
\begin {equation}
 X_t=e^{{\sqrt \alpha} B_t+{{(\mu - \alpha)\over 2}{t }}}.
\end{equation}
This is the optimal structure of protein structure.$^1$ It will be worth to note that eq.(1.6) includes perturbed drift as a result there is an underlying displacement.

\subsection{Trading Strategy}
\par
Trading strategy is a concept in mathematical finance given by a twin of stochastic variables denoted by$^2$
 and provides a powerful tool in biology which is
 \begin {equation}
 (a_t,b_t).
\end{equation}
In this equation $a_t$ and $b_t$ are given by as defined in the space of square integral function $L^2(R^n)$
\begin {equation}
a_tdX_t=\mu X_tdt +{ \sqrt\alpha}X_tdB_t
\end{equation}
and
\begin {equation}
db_t=rb_t\beta_t dt, 
\end{equation}
where $r$ denotes interest ratio and $\beta_t$ is mean gain of entropy process of protein given by
\begin {equation}
d\beta_t=rb_t\beta_t dt.
\end{equation}
Using these notation of the portfolio, which is a twin of stochastic variable, is given by
 \begin {equation}
V_t=a_tX_t+b_t\beta_t.
\end{equation}
This provide the degree of free energy loss.

\subsection{Adaptive Optimization}
\par
Let us consider a stochastic integral of optimal pricing  w.r.t.  trading strategy as follows
\begin {equation}
 X_t=\int^t_0 X_s d\{a_s,b_s\}_s .
\end{equation}
If r.h.s. does not change pricing is called adoptive to trading strategy. This defines a  martingale.$^3$
This means that protein structure is globally optimized to trading strategy. This equation can be solved by iteration as follows.
\par 
{\hskip 1 in \bf repeat until}
\begin {equation}
{\bf E(X_{m}|\{a_n,b_n\}_n)= X_n < \epsilon,  \, a.e. \, \,  m>n}
\end{equation}
\par
{\hskip 1 in \bf continue}
\par
This provides an algorithm implementable into digital computer where  the optimal structure adoptive to trading strategy in which l.h.s denotes a conditional probabilty. In this case, trading strategy needs not be self-balancing. It will be hoped to contribute protein folding whose importance is well recognized.$^4$

\section{Concluding remarks}
\par
Optimal folding for protein structure adoptive to trading strategy was formulated in terms of Ito type stochastic differential equation. It should be noted, however, that there is a limitation in present formulation that it rely on Brownian motion. As a result it depends on Gaussian process based on central limit theorem which states summation of independent variable is governed by Gaussian distribution. The shortcomings arises from the fact that variables are not independent in the real world. An attempt is under progress to incorporate correlation.



\bibliographystyle{amsplain}

\begin{thebibliography}{10}

\bibitem {1} T.Mikosh, \textit{Mechanisms of Protein Folding}  (Oxford University Press 2000).

\bibitem {2} D.Duffie, \textit{Dynamic Asset Pricing Theory} (Princeton University Press 1996).

\bibitem {3} T.Hida, \textit{Brownian Motion} (Springer-Verlag 1980).

\bibitem {3} D.Applebaum, \textit{Levy Processes and Stochastic Calculus} (Cambridge Univesity Press 2004  ).

\bibitem {4} D.Applebaum, \textit{Levy Processes and Stochastic Calculus} (Cambridge Univesity Press 2004  ).R.H.Pain ed., {\it Mechanisms of Protein Folding}, Springer-Verlag (2000) {\it Original version is ref. 1)}.

\end{thebibliography}

\end{document}